\documentclass[aps,prl, reprint,superscriptaddress,showpacs,amsmath
,amssymb]{revtex4-1}

\usepackage{graphics}
\usepackage{mathtools}
\usepackage{graphicx}% Include figure files
\usepackage[version=3]{mhchem} % Formula subscripts using \ce{}
\usepackage[T1]{fontenc}       % Use modern font encodings
\usepackage{color}
\usepackage{MnSymbol}
\usepackage{tikz}
\usepackage{amsfonts}
\usepackage{amsmath}
\begin{document} 

\title{Measuring geometric phase without interferometry}

\author{T. Malhotra}
\affiliation{Department of Physics, University of Rochester, Rochester, New York 14627, USA}
\affiliation{Center for Coherence and Quantum Optics, University of Rochester, Rochester, NY 14627, USA}
\author{R. Guti\'{e}rrez-Cuevas}
\affiliation{Center for Coherence and Quantum Optics, University of Rochester, Rochester, NY 14627, USA}
\affiliation{The Institute of Optics, University of Rochester, Rochester, NY 14627, USA}
\author{J. Hassett}
\affiliation{Center for Coherence and Quantum Optics, University of Rochester, Rochester, NY 14627, USA}
\affiliation{The Institute of Optics, University of Rochester, Rochester, NY 14627, USA}
\author{M. R. Dennis}
\affiliation{H. H. Wills Physics Laboratory, University of Bristol, Bristol BS8 1TL, UK}
\affiliation{School of Physics and Astronomy, University of Birmingham, Birmingham B15 2TT, UK}
\author{A. N. Vamivakas}
\email{nick.vamivakas@rochester.edu}
\affiliation{Department of Physics, University of Rochester, Rochester, New York 14627, USA}
\affiliation{Center for Coherence and Quantum Optics, University of Rochester, Rochester, NY 14627, USA}
\affiliation{The Institute of Optics, University of Rochester, Rochester, NY 14627, USA}
\affiliation{Materials Science, University of Rochester, Rochester, New York 14627, USA}
\author{M. A. Alonso}
\email{miguel.alonso@rochester.edu}
\affiliation{Center for Coherence and Quantum Optics, University of Rochester, Rochester, NY 14627, USA}
\affiliation{The Institute of Optics, University of Rochester, Rochester, NY 14627, USA}
\affiliation{Aix Marseille Universit\'e, Centrale Marseille, Institut Fresnel, UMR 7249, 13397 Marseille Cedex 20, France}

\begin{abstract}
A simple non-interferometric approach for probing the geometric phase of a structured Gaussian beam is proposed. Both the Gouy and Pancharatnam-Berry phases can be determined from the intensity distribution following a mode transformation if a part of the beam is covered at the initial plane. Moreover, the trajectories described by the centroid of the resulting intensity distributions following these transformations resemble those of ray optics, revealing an optical analogue of Ehrenfest's theorem associated with changes in geometric phase.
\end{abstract}

\maketitle

In 1984, almost 30 years after Pancharatnam \cite{pancharatnam1956generalized} first noticed a geometric phase in light polarization, Berry \cite{berry1984quantal,berry1987adiabatic} discovered that quantum systems acquire not only a dynamic phase due to time evolution but also a geometric phase dependent on the path taken in their parameter space. This geometric phase, known also as the Pancharatnam-Berry (PB) phase, depends only on the parameter space's geometry. Geometric phases have undergone extensive generalizations, led to many applications \cite{anandan1997resource,wilczek1989geometric,galvez2002applications,
calvo2005wigner,
tamate2009geometrical,milione2015using,slussarenko2016guiding,yale2016optical}, and become a unifying concept in physics. 
Geometric phases of light appear in many scenarios, such as polarization \cite{pancharatnam1956generalized} and changes in propagation direction \cite{tomita1986observation}.  Changes in the transverse modal structure of an optical beam can also lead to a geometric phase \cite{van1993geometric,padgett1999poincare,habraken2010universal,galvez1999geometric,galvez2003geometric,milione2012higher} understood in the context of a spatial mode Poincar\'e sphere. It is this last type of geometric phase that is the main focus of this work.
\begin{figure}
\centering
\includegraphics[width=1\linewidth]{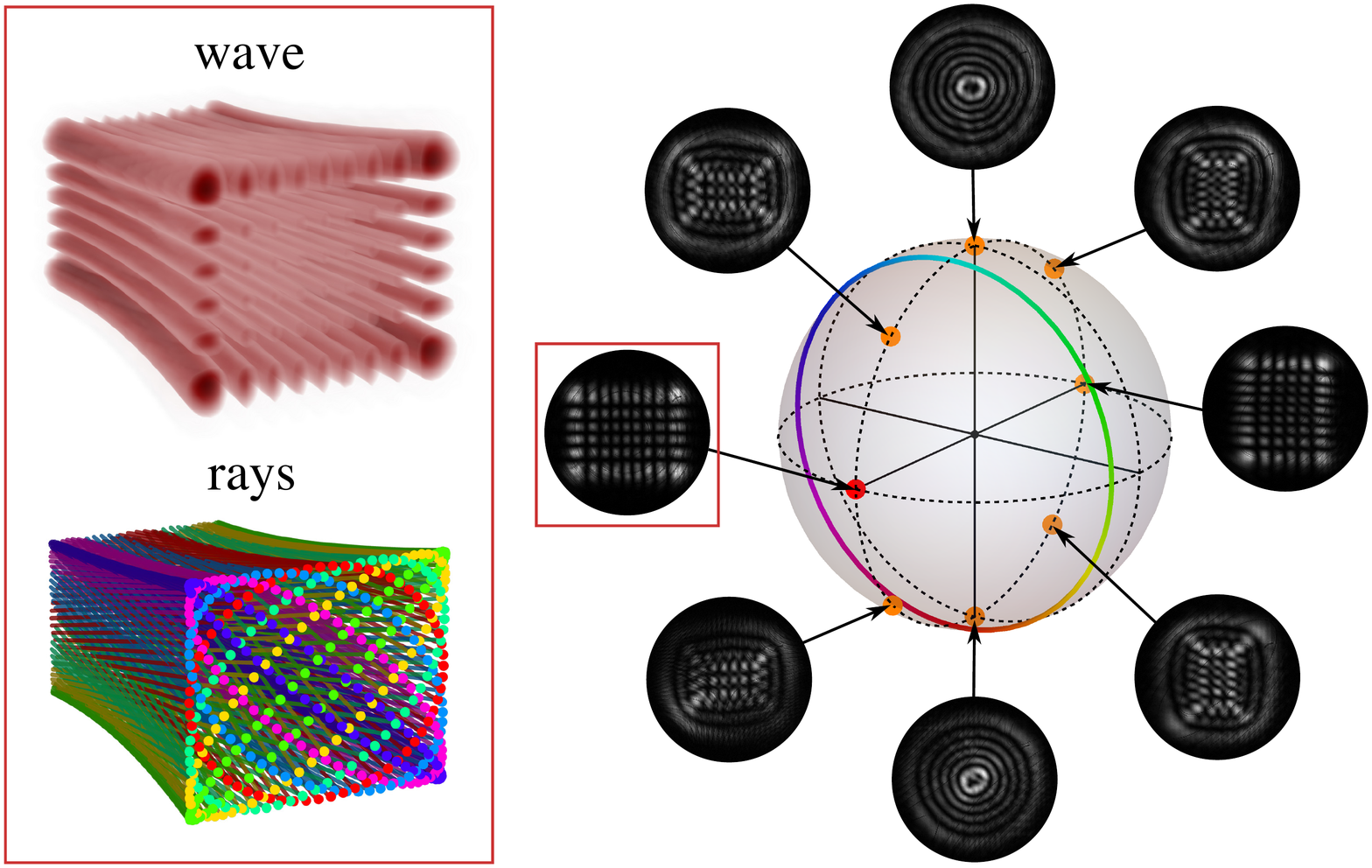}
\caption{\label{fig:rps} Modal Poincar\'e sphere for $p=5$, $l=3$. The red frame identifies the input horizontal HG$_{8,5}$ mode used in the experiments. The colored circle over the sphere
corresponds to the Poincar\'e path (PP) of this mode, where each point corresponds to a family of rays whose cross section is an ellipse. Some of these ellipses of rays are shown in the bottom-left picture labeled ``rays'', where color is used to identify these ellipses with points along the PP. These colors represent different values of $\eta$, while different rays within each ellipse correspond to different values of $\tau$. The 3D picture of the beam's wave intensity is shown above that for the rays.
The center of the PP is the modal spot, shown as a red dot.  Also shown are the experimental intensity distributions of other modes, corresponding to the modal spots shown as orange dots. (The corresponding PPs and ray distributions are not shown.) %The figures to the left represent in 3D the intensity (top) and the ray construction (bottom) for the HG$_{8,5}$ mode.  
}
\end{figure}

The modal Poincar\'e sphere (MPS) was proposed for first-order structured Gaussian beams \cite{padgett1999poincare} where the two poles correspond to Laguerre-Gauss (LG) modes with equal circular shape but opposite vorticity. All other points over the sphere correspond to complex linear combinations of these two modes. In particular, points along the equator 
correspond to rotated Hermite-Gauss (HG) modes with Cartesian orders $m=1,n=0$. This MPS construction was later extended  \cite{habraken2010universal} to characterize higher-order modes. The two poles are again assigned to LG modes with equal radial order $p$ and azimuthal orders $\pm l$ ($l \geq 0$). The rest of the sphere corresponds not to linear combinations of these two modes but to the generalized Hermite-Laguerre-Gauss (HLG) modes \cite{abramochkin2004generalized}, with points over the equator corresponding again to rotated HG modes with Cartesian orders $m=p+l,n=p$. Figure~\ref{fig:rps} shows an example of this MPS corresponding to $p=5,l=3$, with experimentally measured modes decorating the sphere's surface.  Given this construction, it is natural that a PB phase arises from a series of optical transformations that traces a closed path over the MPS.  
  
\begin{figure*}
\centering
\includegraphics[width=1\linewidth]{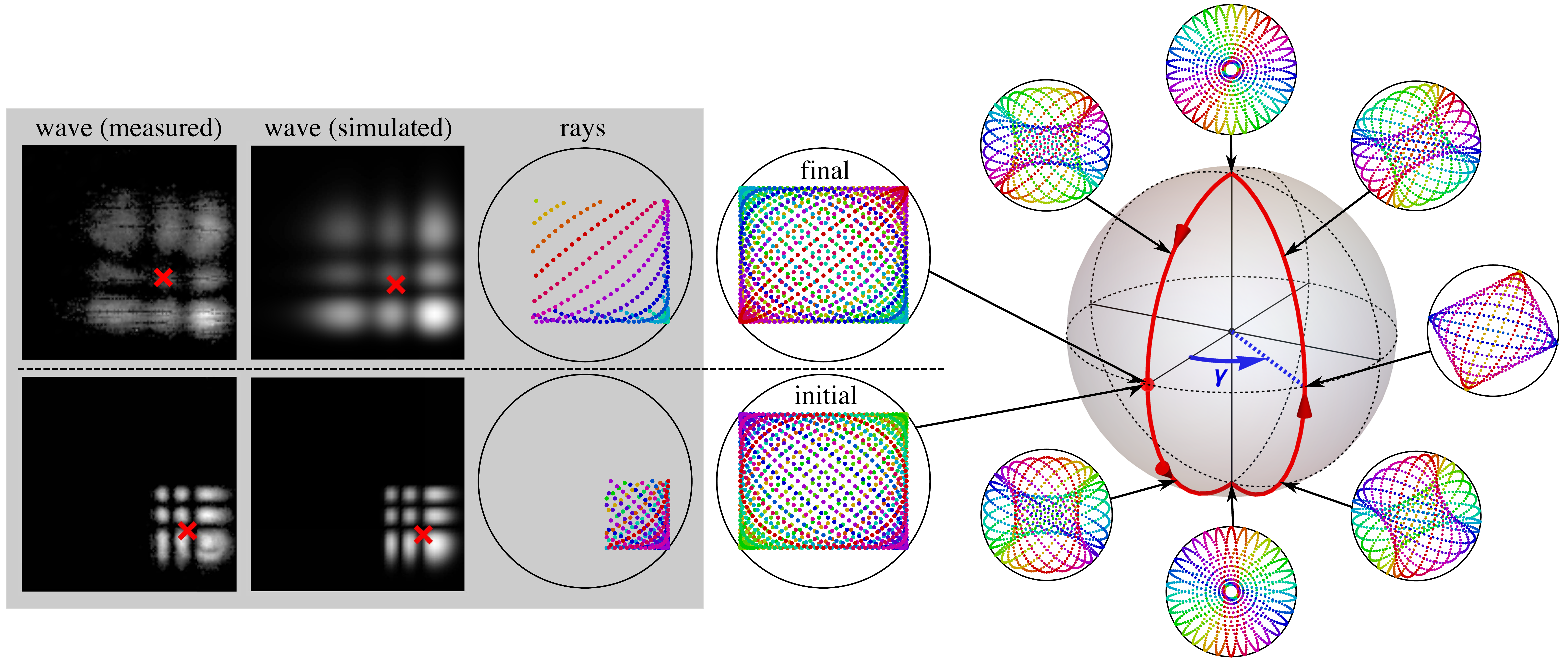}
\caption{\label{fig:path} 
Mode transformation corresponding to the closed (red) path traced by the modal spot over the MPS. The ray distributions at different stages along the path are shown inside the circles around the MPS, where points of equal color correspond to equal values of $\eta$. 
The initial and final ray distributions (directly to the left of the MPS) have the same rectangular shape and correspond to the same HG mode, but the different color distribution reveals the cycling of the rays that gives rise to the PB phase. 
For the initial ray distribution the colors of the ellipses identify points of the PP in Fig.~\ref{fig:rps}. 
The figures on the left (gray background) show the effect of blocking all but the lower right corner of the initial HG beam, before (bottom row) and after (top row) the mode transformation, and the resulting drift in intensity centroid (red crosses) for the measured and simulated intensities, and the transmitted rays. 
}
\end{figure*}
Here we present a simple noninterferometric approach for measuring the PB phase. This approach 
emerges from a deeper understanding of structured Gaussian beams, based on an intuitive ray model \cite{dennis2017swings,alonso2017ray}
that explains both the PB and Gouy phases. Structured Gaussian beams can be described in terms of a ray family in which each ray is specified by the values of two periodic parameters, $\tau$ and $\eta$. 
%Mathematical details of this parametrization are given in the Supplemental Materials, so here we only give a qualitative description. 
At any transverse plane, the rays corresponding to all values of $\tau$, for fixed $\eta$, trace an ellipse with given orientation, handedness and eccentricity (see Supplemental Materials for details). In analogy with polarization, this ellipse of rays corresponds to a point on the Poincar\'e sphere.  The second variable, $\eta$, parametrizes a closed loop over the sphere, referred to here as the Poincar\'e path (PP), shown as a colored circle in Fig.~\ref{fig:rps}, for a HG mode. Each beam corresponds not to a point but to an extended path over the sphere. The transverse ray structure is then a continuous superposition of ellipses, where each point of each ellipse is a ray. For HG, LG and more general HLG modes, the PP is simply a circle, whose center (also shown in Fig.~\ref{fig:rps}) corresponds to the spot used in the standard MPS representations  \cite{padgett1999poincare,agarwal19992,calvo2005wigner,
habraken2010universal}, so we call it the modal spot. Note that the ray elipses form envelopes, i.e. caustics, in the vicinity of which the main intensity features are localized. For HG modes these caustics have a rectangular shape (see Fig.~\ref{fig:rps}), consistent with the fact that the wave solution is separable in Cartesian coordinates. Similarly, for a LG mode, separable in polar coordinates, the caustics are two concentric circles. Note also that modes with equal total mode order $N$ belong in the same sphere but correspond to different PPs. For example, the PPs for HG modes with equal $N=m+n$  are circles centered at the same modal spot but with different sizes, enclosing solid angle quantized as $2\pi(2n+1)/(N+1)$ \cite{alonso2017ray}.

Conversion between modes is possible using a series of anisotropic quadratic phase masks implemented with spatial light modulators (SLMs)  \cite{rodrigo2006optical,alieva2007orthonormal}. These transformations have the effect of rotating the MPS around an axis within the equatorial plane. The orientation of this axis depends on the orientation of the quadratic phases, and the angle of rotation depends on their strength. A sequence of transformations can be considered that brings the modal spot of the circular PP back to its initial position.  One such example is presented in Fig.~\ref{fig:path}, where the modal spot starts and ends at an equatorial point that corresponds to a HG mode. Even though the final and initial modal spots coincide, each point of the PP is shifted according to $\eta\to\eta-\Omega$, where $\Omega=2\gamma$ is the solid angle traced by the modal spot, and $\gamma$ is the angle between the segments of the trajectory. This transformation sequence is equivalent to a single rotation by $\Omega$ of the sphere around the direction of the initial/final modal spot. 
Wave-optically, this rotation corresponds to an anamorphic fractional Fourier transform (see Supplemental Materials) acting on the mode, which can be written in operator form as
\begin{align}
\label{eq:affh}
\exp\left[{\rm i}\,\gamma\left(\hat{H}_x-\hat{H_y}\right)\right],
\end{align}
where $\hat{H}_q$ for $q=x,y$ has the form of a 1D harmonic oscillator Hamiltonian given by
\begin{align}
\hat{H}_q=\frac{q^2}{w_0^2}-\frac{w_0^2}4\frac{\partial^2}{\partial q^2},
\end{align}
with $w_0$ being the waist width \cite{agarwal1994simple}.
The modal PB phase is then derived using an operator formalism analogous to that used in quantum mechanics  \cite{enk1992eigenfunction,dennis2017swings}. Since HG modes are eigenstates of the operator $\hat{H}_x-\hat{H_y}$ with eigenvalue $N-2n$ the effect of the operator in Eq.~(\ref{eq:affh}) is then to produce the PG phase $\Phi=(N-2n)\gamma$. %, which is exactly the PB phase acquired along the path shown in Fig.~\ref{fig:path}. 
Within the ray picture, this transformation has an intuitive geometric interpretation: it simply corresponds to a cycling of the roles played by the different ellipses of rays (compare the initial and final ray configurations in Fig.~\ref{fig:path}). The PB phase is then caused by a change in the role that each ray plays in the beam profile. 

\begin{figure}
\centering
\includegraphics[width=1.\linewidth]{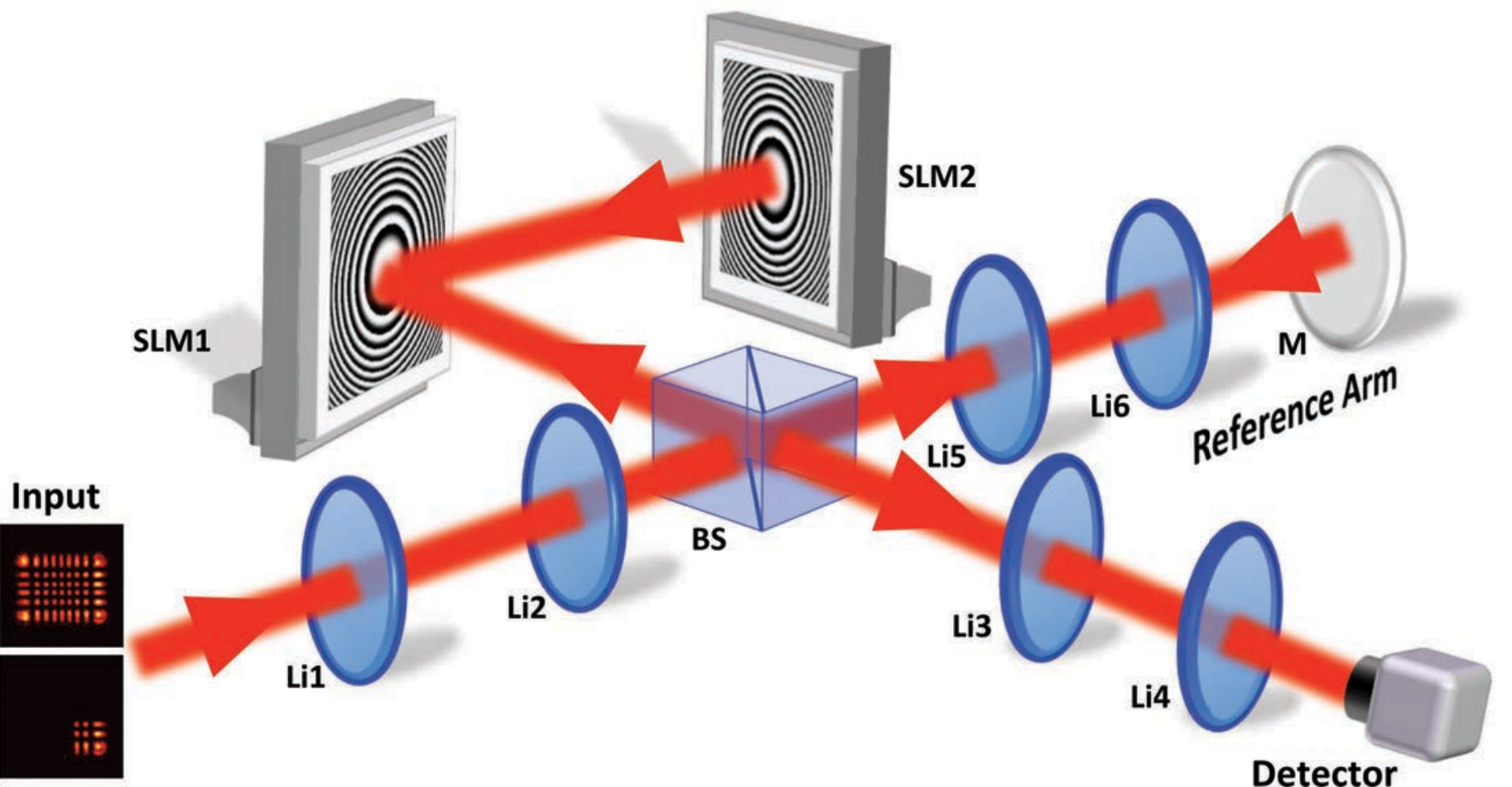}
\caption[Optical set-up]{
Schematic of the optical set-up. For the non-interferometric measurement, only the arm of the Michelson interferometer that contains the SLMs is used. The anisotropic lenses L1 and L3 are implemented on SLM1, while L2 is implemented on SLM2.  The input field is imaged onto SLM1 using a 4f system (unit magnification) formed by lenses Li1 and Li2. In the reference arm (used for the interferometric measurements), the field is relayed onto mirror M using another telescope formed by lenses Li5 and Li6 (magnification=0.5). The interference signal is relayed to the CCD detector using the 4f system formed by lenses Li3 and Li4 (magnification=0.5). Distances are not drawn to scale.
}
\label{fig:setup}
\end{figure}

The proposed method for probing the geometric phase exploits this cycling. If part of the initial beam is occluded, the shadow in the final beam is at a different part of its transverse profile, its location linked to the acquired PB phase.  Figure \ref{fig:path} shows the effects of this occlusion for both rays and waves, where only the lower-right corner of the initial HG beam is unblocked. Following the modal transformation, the unblocked rays spread out and drift towards the upper-left corner due to the ray cycling caused by the transformation. The simulated and experimentally measured wave intensities exhibit the behavior anticipated by analyzing the rays. The position of the intensity distribution centroid is sufficient for determining the PB phase. The intensity centroid $(x_c,y_c)$ after the transformation is just a linear combination of the intensity centroid $(x_0,y_0)$ of the initial beam and that of its Fourier transform \cite{bastiaans2005wigner,healy2015linear}. Further, because the Fourier-space centroid vanishes when the initial beam is a blocked HG mode, the centroid for the transformation considered here is simply
\begin{align}
\label{eq:cent}
(x_c,y_c)=&(x_0,y_0) \cos \gamma.
\end{align}
This centroid gives access to $\gamma$ within the range $[0,\pi)$, from which the PB phase can be deduced. For $N-2n=l>2$ this allows discriminating between different multiples of $2\pi$, in contrast with interferometric approaches 
\cite{galvez2003geometric}.
\begin{figure}
\centering
\includegraphics[width=.99\linewidth]{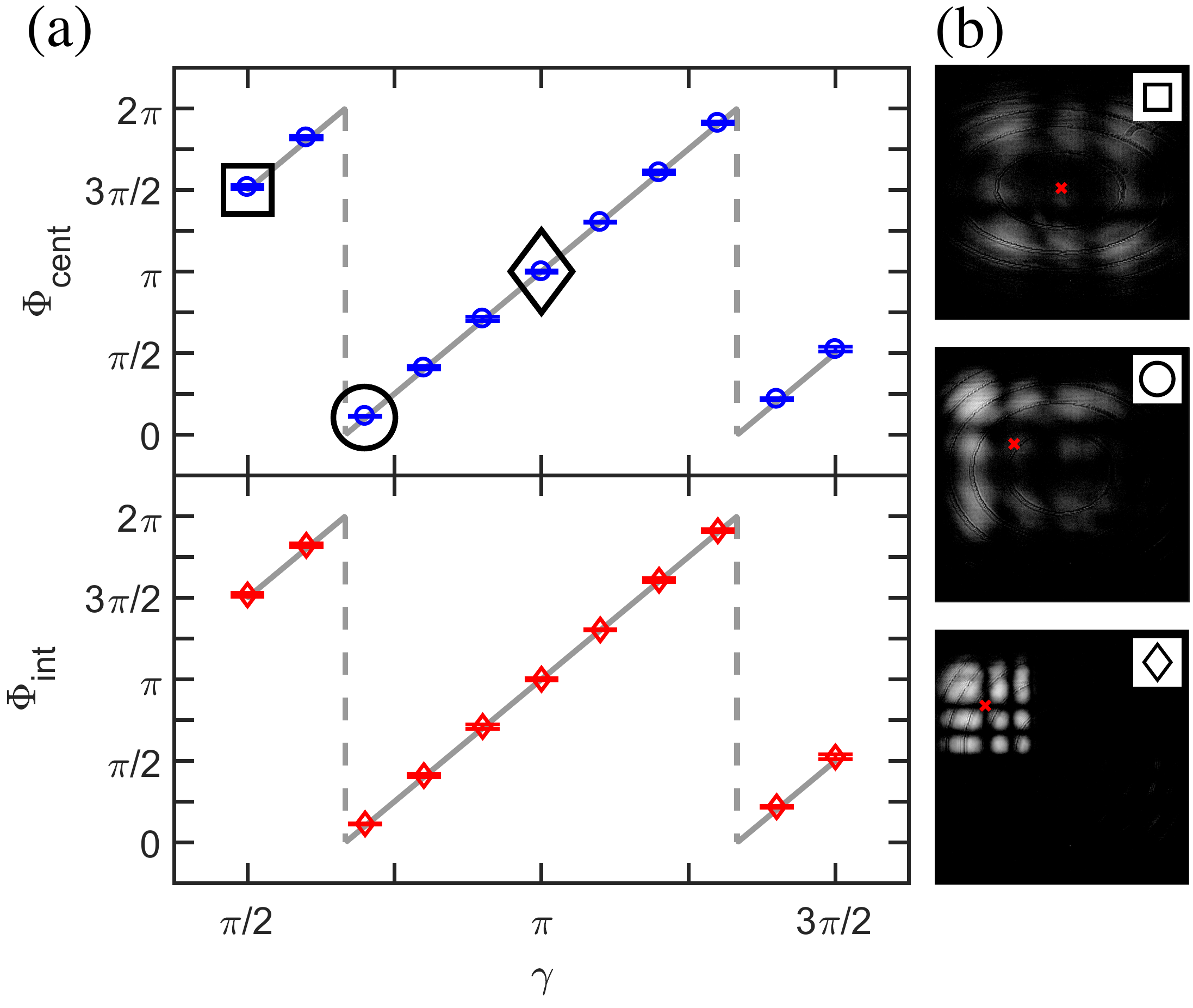}
\caption[Experimental data]{ \linespread{1.1}\selectfont{PB phase measurement results.
 (a) Measured PB phase as a function 
of $\gamma$ using the centroid of the blocked beam (circles, top panel), and 
interferometric measurements (diamonds, bottom panel). The non-interferometric results are wrapped onto $[0,2\pi)$ for comparison with the interferometric ones.  Error bars are shown for both, as well as the theoretical value (gray line). (b) Measured intensity of the
transformed blocked beam (plotted on a log scale) with centroids as red crosses, for different $\gamma$. 
The symbols at the insets correspond to the markers in (a). 
}}
\label{fig:data}
\end{figure}

In our experiments, the mode transformation is based on a set-up \cite{rodrigo2006optical,malhotra2018interferometric} that uses three anisotropic lenses equally separated by a distance $z$, as shown in Fig.~\ref{fig:setup}. The powers of these lenses are parametrized by the angles $\alpha_x, \alpha_y$ as
\begin{subequations}
\label{eq:power}
\begin{align}
p_{q}^{(\text{L1,L3})}&=\left[1-\cot\left(\alpha_{q}/2\right)/2\right]/z , \\
p_{q}^{(\text{L2})}&=2\left(1-\sin\alpha_{q}\right)/z,
\end{align}
\end{subequations}
where L1, L2 and L3 denote the three lenses and $p_{q}^{(\text{L}j)}$ is the power of L$j$ along the $q$ direction, with $q=x,y$. (The relation between the angles $\alpha_x, \alpha_y$ and the system's geometric phase is discussed in the Supplemental Materials.) The lenses are implemented electronically by displaying their phase transmittances on two SLMs controlled using LabView. Note that L2 is implemented in reflection mode, so L1 and L3 correspond to the same SLM. 
The desired input beam (a $\text{HG}_{8,5}$ mode, shown in Fig.~\ref{fig:rps}) is prepared by illuminating a third SLM in a different set-up (not shown in Fig.~\ref{fig:setup}) with a collimated laser beam ($\lambda$ = 795 nm) polarized along the SLM's preferred axis. A metallic mask occludes part of the input beam and the intensity is recorded by a CCD. The intensity of the obscured input beam is shown in the lower-left corner of both Figs.~\ref{fig:path} and \ref{fig:setup}. The centroid coordinates ($x_c, y_c$) are computed from the recorded intensity and Eq.~(\ref{eq:cent}) is used to extract two values for $\gamma$, which are averaged to obtain the final estimate. Notice from Fig.~\ref{fig:setup} that the set-up also includes a reference arm, not used for the centroid measurements.

\begin{figure}
\centering
\includegraphics[width=.9\linewidth]{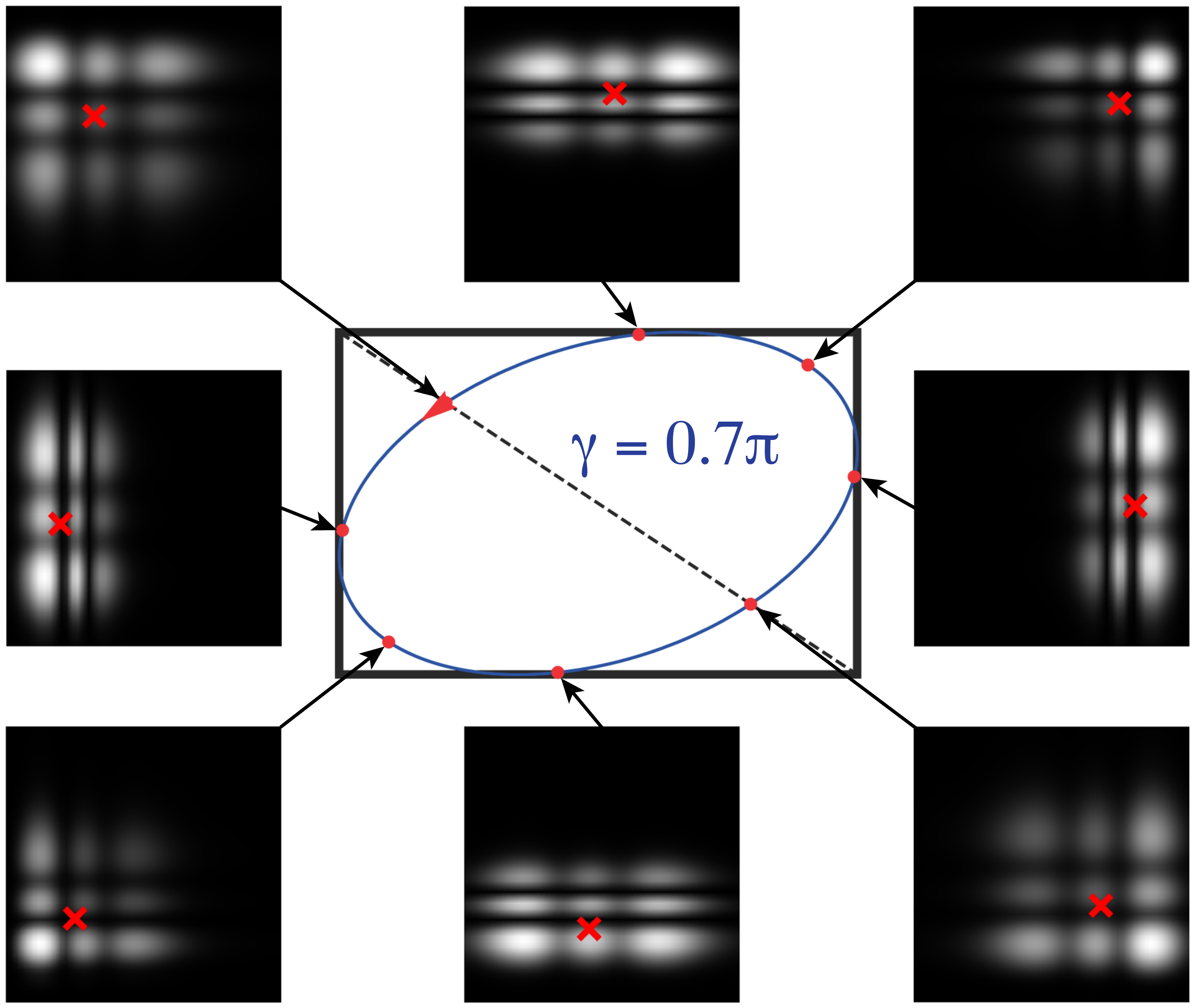}
\caption{\label{fig:ellipse} Evolution of the intensity distribution of a blocked HG beam as the Gouy phase varies, and the corresponding elliptic trajectory traced by the centroid for a fixed PB phase.}
\end{figure}

For validation, interferometric measurements are also performed by sending the complete HG mode through both the test and reference arms (see Supplemental Materials for details). 
%A piston phase $\psi$ (varying between 0 and 2$\pi$) is implemented on SLM2, and the  power  resulting from the interference from both arms is recorded by a bucket detector as a function of $\psi$. The interferometer is path-stabilized (see \cite{malhotra2018interferometric} for details) to ensure that the two arms of the interferometer remain in a fixed path-difference configuration during the data acquisition. We use as a reference an interference measurement resulting from using the set-up in the inverted imaging mode (that is, using $\alpha_x=\alpha_y=\pi$) to deduce the phases corresponding to different values of $\gamma$. The negative branch of the arccosine function was chosen for $\alpha > \pi$ in order to match the measured interferometric PB phase. The interferograms are fitted with sinusoids to recover the phase, and the phase of the reference measurements (which include an extra phase of $\pi$ due to the image inversion) is subtracted, leading to the desired PB phase. 
Figure~\ref{fig:data} shows the agreement between the PB phases obtained via interferometry and the centroid measurements, for multiple values of $\gamma$ corresponding to many paths over the MPS.

\begin{figure}
\centering
\includegraphics[width=.8\linewidth]{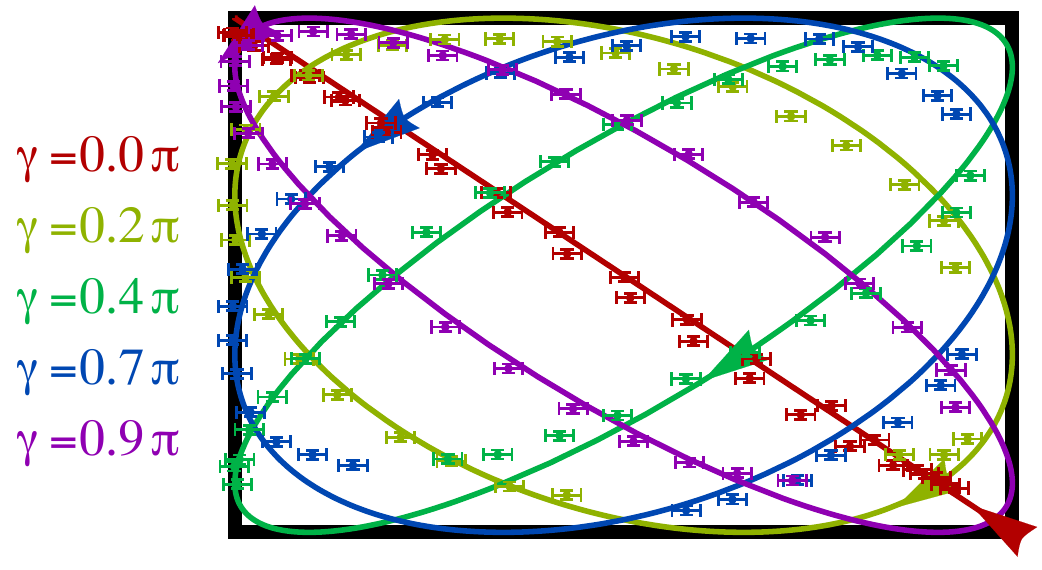}
\caption{\label{fig:ellipses}  Elliptic trajectories traced by the centroid given a constant PB phase determined by the values of $\gamma$ (between $0$ and $0.9\pi$).  The solid curves correspond to the theoretical expectation.  Error bars for centroid determination are included in both directions. The arrowheads indicate the location of zero Gouy phase and the direction of its increase.}
\end{figure}

An important feature of the intensity centroid measurement is its insensitivity to dynamic phase and its ability to determine also the Gouy phase. 
This phase is a result of the increase in spacing between wavefronts near the focal regions of any beam, and for the beams considered here, it constitutes an extra phase of $(N+1)\pi/2$ between the waist plane and the far zone \cite{ozaktas1994fractional,alonso2017ray}. 
Different interpretations have been given to this phase \cite{simon1993bargmann,subbarao1995topological,feng2001physical}, but here we focus on its connection with ray optics. Within the ray picture of structured Gaussian beams used here, the Gouy phase corresponds to a shift in the other ray parameter, $\tau \rightarrow \tau + \zeta$, where $\zeta = \arctan(z/z_{\rm R})$ with $z_{\rm R}$ being the beam's Rayleigh range \cite{ozaktas1994fractional,alonso2017ray}. That is, while the PB phase corresponds to a cycling of the ellipses of rays (a shift in $\eta$), the Gouy phase corresponds to a cycling of rays within each ellipse. This cycling also has the effect of moving the obstruction's shadow, and the resulting intensity centroid (see Fig.~\ref{fig:ellipse}) is now given by
\begin{align}
\label{eq:centGP}
(x_c,y_c)=&[x_0 \cos(\xi +\gamma),y_0\cos(\xi -\gamma)] .
\end{align}
where $\xi=(N+1)\zeta/N$. 
Therefore, both the Gouy and PB phases can be inferred from the centroid position, without the need for diffraction calculations \cite{hamazaki2006direct}. 

An interesting manifestation of Ehrenfest's theorem emerges in this context: the centroid coordinates in Eq.~(\ref{eq:centGP})  mimic the ray positions for the HG modes [compare with Eq.~(S3) in the Supplemental Material], with $\gamma$ playing the role of $\eta$ and $\xi$ that of $\tau$. Like the rays, the centroids are constrained to a rectangle centered at the origin with upper-right corner coordinates $(|x_0|,|y_0|)$. This is shown in Fig.~\ref{fig:ellipses} for both theory and experimental measurements. This rectangular envelope is a scaled version of the caustics for the beam \cite{alonso2017ray}.  
For fixed PB phase ($\gamma$) and varying Gouy phase ($\xi$), the centroid traces an ellipse that is inscribed in this rectangle, similar to the ray ellipses.  

%------------------------------------------------------------------------
%--------------------------------Conclusions-----------------------------
%------------------------------------------------------------------------

In summary, a non-interferometric method for measuring geometric phases in structured Gaussian beams is presented. The approach is motivated by the toroidal structure (involving two periodic parameters) of the ray family associated with these beams, and it relies on the fact that the PB and Gouy phases correspond to shifts on each of the two ray parameters (two different rotations of this torus). These shifts have no effect on the intensity of the unperturbed beam, but they become appreciable when part of the beam is blocked. 
Note that when the blocked part is not too large, we can view this phenomenon as a so-called ``healing'' effect \cite{bouchal1998self,alonso2017ray}, in which the blocked features are restored by the mode transformation at the cost of the shadow moving elsewhere in the beam profile. 
These results highlight the conceptual power of the ray picture as a way to understand the internal structure of the beam, and provide an example of the similarity of the behavior of rays and intensity centroids according to Ehrenfest's theorem, not only for evolution under free propagation but also under the more complex modal transformations considered here. Finally, while we focused on geometric phases for structured beams, variants of this approach can be applied to other incarnations of geometric phase through the observation of the effects of perturbations in the incoming state. For polarization, for example, a dichroic element can be used to modify the initial polarization, whose effect on the output would reveal the geometric phase \cite{wagh1995measuringa,loredo2009measurement}.
\\

T.~M. and A.~N.~V. acknowledge support by ARO W911NF-16-1-0162 and the Leonard Mandel Faculty Fellowship in Quantum Optics. R.~G.-C. was supported by a CONACYT fellowship. M.~A.~A. acknowledges support from the National Science Foundation (PHY-1507278) and the Excellence Initiative of Aix-Marseille University - A*MIDEX, a French ``Investissements d'Avenir'' programme. M.~R.~D. was supported by the Leverhulme Trust Research Programme Grant No. RP2013-K-009, SPOCK: Scientific Properties of Complex Knots. 
T.~M. and R.~G.-C. contributed equally to this work.

%\bibliography{refs}

%

\end{document}